\documentclass[]{spie}  
\usepackage{graphicx,amsmath,amssymb,latexsym}

\title{Co-phasing the Large Binocular Telescope: status and performance of LBTI/PHASECam}

\author{D.~Defr\`ere\supit{a}, P.~Hinz\supit{a}, E.~Downey\supit{a},  D.~Ashby\supit{b},  V.~Bailey\supit{a},  G.~Brusa\supit{a}, J.~Christou\supit{b}, W.C.~Danchi\supit{c},  P.~Grenz\supit{a}, J.M.~Hill\supit{b}, W.F.~Hoffmann\supit{a}, J.~Leisenring\supit{a}, J.~Lozi\supit{a}, T.~McMahon\supit{a}, B.~Mennesson\supit{d}, R.~Millan-Gabet\supit{e},  M.~Montoya\supit{a}, K.~Powell\supit{a}, A.~Skemer\supit{a},  V.~Vaitheeswaran\supit{a}, A.~Vaz\supit{a}, and C.~Veillet\supit{b}  
\skiplinehalf
\supit{a}Steward Observatory, University of Arizona, 933 N. Cherry Avenue, 85721 Tucson, USA\\
\supit{b}Large Binocular Telescope Observatory, University of Arizona, 933 N. Cherry Avenue, 85721 Tucson, USA\\
\supit{c}NASA Goddard Space Flight Center, Exoplanets \& Stellar Astrophysics Laboratory, Code 685, Greenbelt, MD 20771\\
\supit{d}Jet Propulsion Laboratory, California Institute of Technology 4800 Oak Grove Drive, Pasadena CA 91109-8099, USA\\
\supit{e}NASA Exoplanet Science Center (NExSci), California Institute of Technology, 770 South Wilson Avenue, Pasadena CA 91125, USA\\
}

\authorinfo{Send email correspondence to D.D. at ddefrere@email.arizona.edu.}

\begin{document} 
\maketitle 

\begin{abstract}
The Large Binocular Telescope Interferometer is a NASA-funded nulling and imaging instrument designed to coherently combine the two 8.4-m primary mirrors of the LBT for high-sensitivity, high-contrast, and high-resolution infrared imaging (1.5-13\,$\mu$m). PHASECam is LBTI's near-infrared camera used to measure tip-tilt and phase variations between the two AO-corrected apertures and provide high-angular resolution observations. We report on the status of the system and describe its on-sky performance measured during the first semester of 2014. With a spatial resolution equivalent to that of a 22.8-meter telescope and the light-gathering power of single 11.8-meter mirror, the co-phased LBT can be considered to be a forerunner of the next-generation extremely large telescopes (ELT).
\end{abstract}

\section{INTRODUCTION}
\label{sec:intro}  

The Large Binocular Telescope\cite{Hill:2014,Veillet:2014} is a two 8.4-m aperture optical instrument installed on Mount Graham in southeastern Arizona (at an elevation of 3192 meters) and operated by an international collaboration between institutions in the United States, Italy, and Germany. Both telescopes are equipped with state-of-the-art high-performance adaptive optics systems that work reliably for science observations \cite{Bailey:2014,Christou:2014}. The Large Binocular Telescope Interferometer (LBTI) is a NASA-funded nulling and imaging instrument designed to coherently combine the two primary mirrors of the LBT for high-sensitivity, high-contrast, and high-resolution infrared imaging (1.5-13\,$\mu$m). It is developed and operated by the University of Arizona and based on the heritage of the Bracewell Infrared Nulling Cryostat (BLINC) on the MMT\cite{Hinz:2000}. It is equipped with two scientific cameras: LMIRCam\cite{Wilson:2008} (the L and M Infrared Camera) and NOMIC\cite{Hoffmann:2014} (Nulling Optimized Mid-Infrared Camera) covering respectively the 3-5$\mu$m and 8-13$\mu$m wavelength ranges. The main scientific goals are to determine the brightness and prevalence of exozodiacal dust and image giant planets around nearby main-sequence stars. Two surveys are currently being carried out in that respect: an exozodiacal dust survey called HOSTS (Hunt for Observable Signatures of Terrestrial Planetary Systems)\cite{Danchi:2014} and a planet survey called LEECH (LBTI Exozodi Exoplanet Common Hunt) \cite{Skemer:2014}. 

\begin{figure}[!t]
	\begin{center}
		\includegraphics[width=14.5 cm]{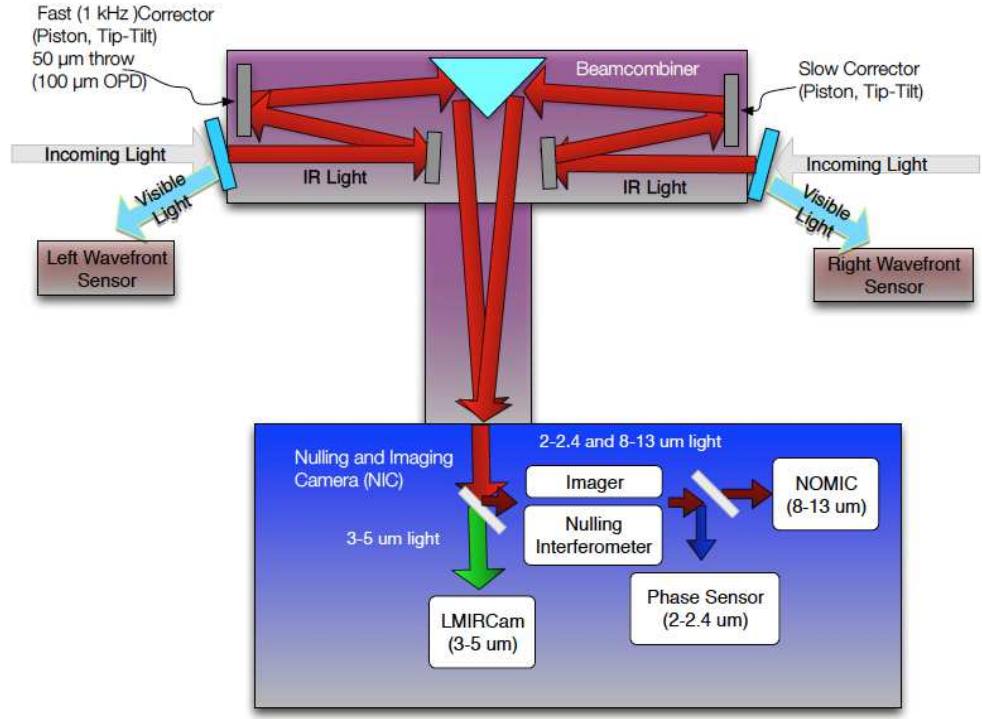}
		\caption{Components of the LBTI shown with the optical path through the beam combiner and the NIC cryostat. Starlight is reflected on LBT primaries, secondaries, and tertiaries before coming into this diagram on the top right and top left. The visible light is reflected on the entrance window and used for adaptive optics while the infrared light is transmitted into LBTI, where all subsequent optics are cryogenic. The beam combiner directs the light with steerable mirrors and can adjust pathlength for interferometry.  Inside the NIC cryostat, 3-5$\mu$m light is directed to LMIRCam for exoplanet imaging, 2.0-2.4$\mu$m light is directed to the phase sensor, which measures the differential tip/tilt and phase between the two primary mirrors, and 8-13$\mu$m light is directed to NOMIC for Fizeau imaging or nulling interferometry.}\label{fig:diagram}
	\end{center}
\end{figure}

The LBT is an ideal platform for interferometric observations because both telescopes are installed on a single steerable mount. This design does not require long delay lines and contains relatively few warm optical elements which provides an exceptional sensitivity. The overall LBTI system architecture and performance are presented elsewhere in these proceedings\cite{Hinz:2014}. In brief,  the LBTI consists of a universal beam combiner (UBC) located at the bent center Gregorian focal station and a cryogenic Nulling Infrared Camera (NIC). The UBC provides a combined focal plane from the two LBT apertures while the precise overlapping of the beams is done in the NIC cryostat (see optical path through the UBC and NIC in Figure~\ref{fig:diagram}). Interferometric combination can be performed in the image plane for Fizeau imaging or in the pupil-plane for nulling interferometry\cite{Defrere:2014c}. Fizeau imaging provides a spatial resolution equivalent to that of a 22.8-meter telescope along the horizontal/azimuthal axis and the light-gathering power of single 11.8-meter mirror. Field rotation can be used to access a range of parallactic angles (PA) and recover the diffraction-limited spatial resolution of a 22.8-meter circular aperture in post-processing. This capability has been recently demonstrated on Jupiter's moon Io \cite{Leisenring:2014}. Employing ``lucky Fizeau'' imaging at M-Band, approximately sixteen independent sources, corresponding to known hot spots on the surface of Io and inaccessible to resolutions of 8-meter class telescopes, have been recovered after image reconstruction of the Fizeau interferometric observations. Warm exozodiacal dust has also been resolved around $\eta$~Crv using fringe-tracked nulling interferometric observations \cite{Danchi:2014}. 

The focus of this paper is to describe the status and performance of PHASECam, LBTI's fast-readout near-infrared camera used to measure tip-tilt and phase variations between the two LBT apertures. We describe  PHASECam and the overall control approach in Section~\ref{sec:phasecam}, the current on-sky performance in Section~\ref{sec:perfo}, and the future work in Section~\ref{sec:prospects}.

\section{PHASECam}\label{sec:phasecam}
\subsection{Overview}

\begin{figure}[!b]
	\begin{center}
		\includegraphics[width=14.5 cm]{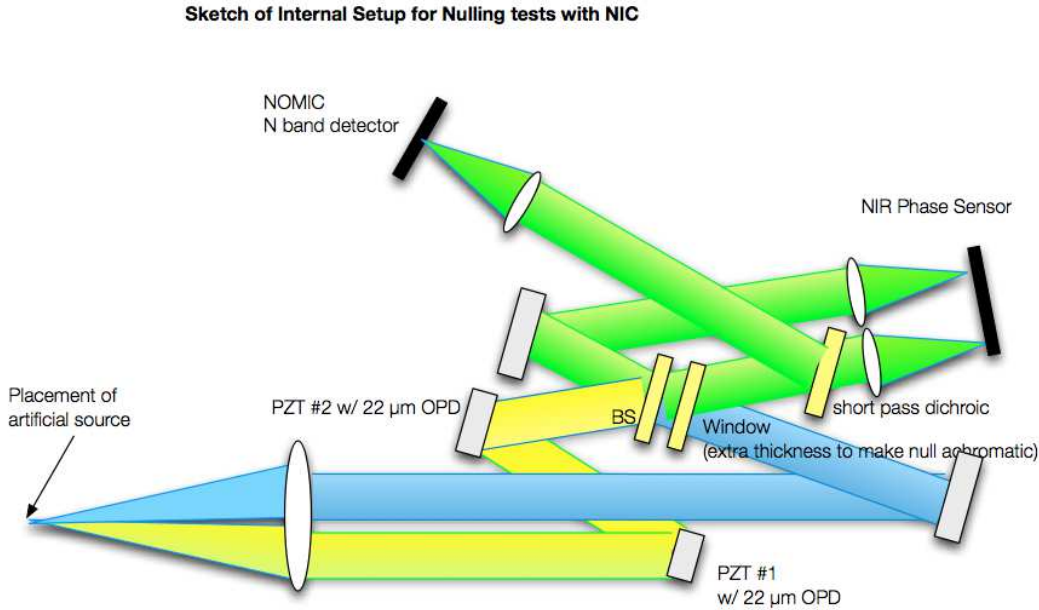}
		\caption{Sketch of the nulling portion of NIC. Both outputs of the interferometer are directed to the near-infrared phase sensor (PHASECam) while one output is reflected to the NOMIC science detector with a short pass dichroic. To provide a flexible approach to phase sensing, the lenses in front of PHASECam can be selected to image either the image plane or the pupil plane. The default approach as shown in Figure~\ref{fig:approach} uses the pupil plane. This particular configuration represents a testing with the internal artificial source located in the image plane on the left side of the figure. }\label{fig:phasing}
	\end{center}
\end{figure}

PHASECam uses a fast-readout PICNIC detector to measure tip/tilt and phase variations between the two AO-corrected LBT apertures. The optical system is designed so that PHASECam receives the near-infrared light  from both interferometric outputs when the long wavelength channel is in either the nulling or the Fizeau imaging mode. In nulling mode, one output of the interferometer is reflected to the NOMIC science detector with a short pass dichroic (see Figure~\ref{fig:phasing}) while, in Fizeau imaging mode, both beams are intercepted before beam combination. The optics provides a field of view of 10~arcsecs with pixels of  0.078~arcsecs wide and can be adapted to create different setups for pathlength sensing. Three options are currently built into the LBTI to allow a flexible approach to phase sensing: 1) use the relative intensity between the two interferometric outputs, 2) use dispersed fringes via a low-dispersion prism, or 3) use an image of the combined pupils via a reimaging lens. Various neutral density filters are also available together with standard H and K filters. 

The LBTI has also two artificial point sources that can be used for phase sensing tests. The first one is a small NiChrome wire within the NIC to test the nulling interferometer and phase sensor, as shown in Figure~\ref{fig:phasing}. The second is a superluminescent diode source at 1.55\,$\mu$m located at the entrance to the beam combiner and can be used to do an end-to-end test of the PHASECam phase sensor. Tip/tilt and pathlength corrections are sent to the Fast Pathlength Corrector (FPC) located in the left part of the beam combiner (Figure~\ref{fig:diagram}). The FPC provides a Piezo-electric transducer (PZT) fast pathlength correction with a 80\,$\mu$m of physical stroke, capable of introducing 160\,$\mu$m of optical path difference (OPD) correction. The right mirror provides a larger stroke (40 mm of motion) for slow pathlength correction (SPC). In practice, the SPC is used to acquire the fringes while the FPC is used to correct atmospheric variations. For internal tests with the NiChrome wire, precise pathlength adjustments can made with two PZT devices (see Figure~\ref{fig:phasing}).  

\subsection{Tip/tilt and phase delay}

\begin{figure}[!t]
	\begin{center}
		\includegraphics[height=5.5 cm]{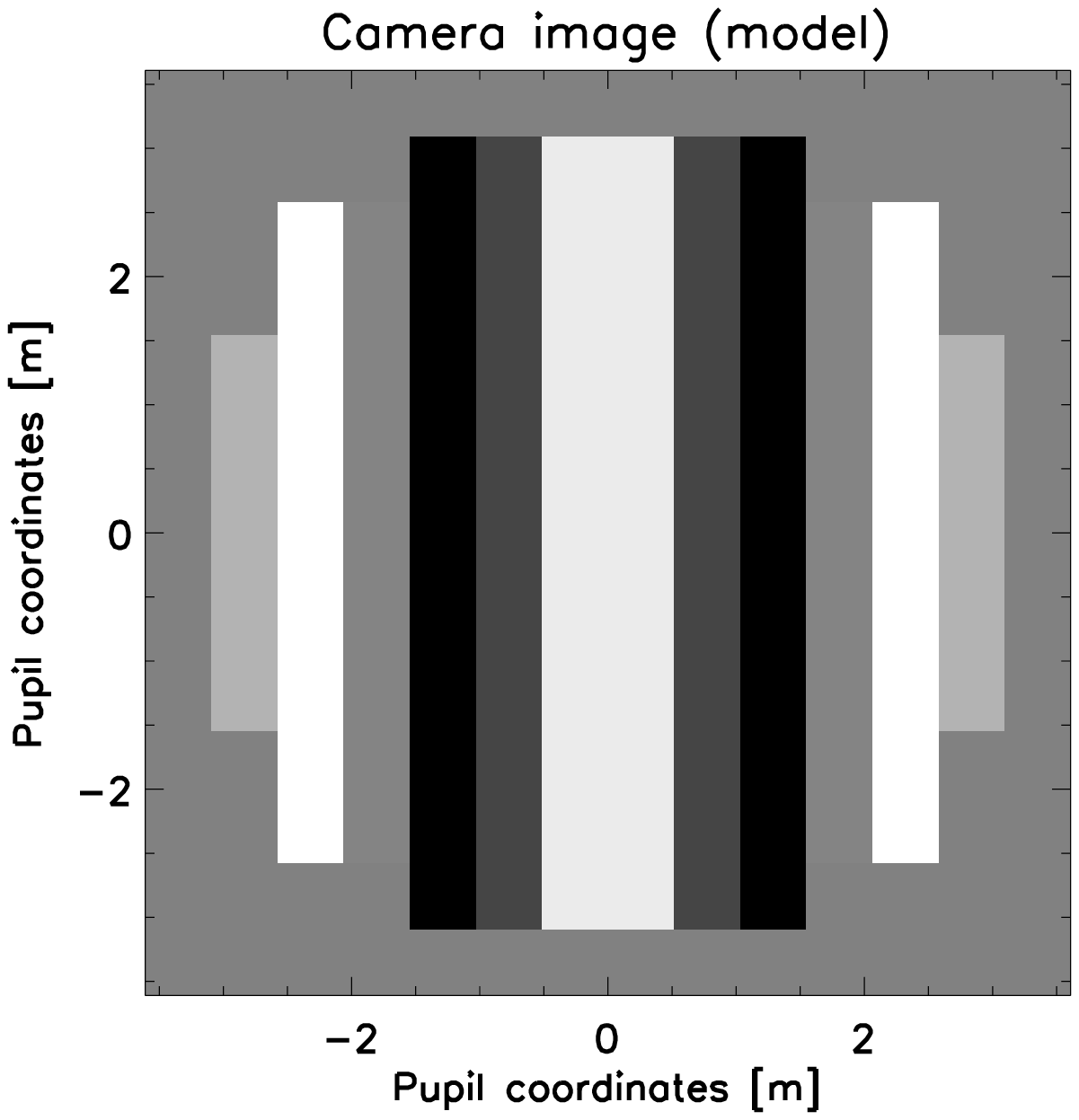}
		\includegraphics[height=5.5 cm]{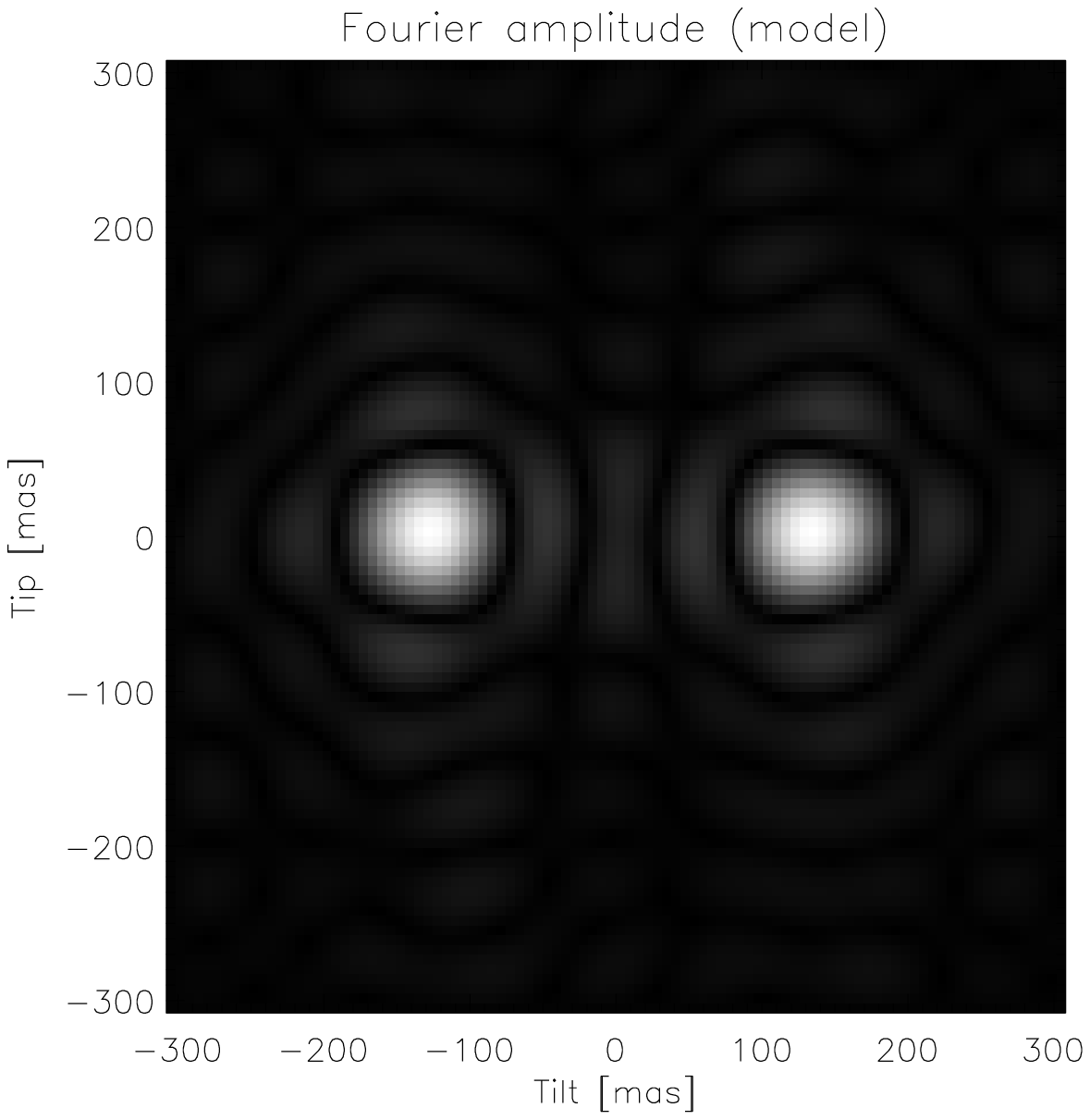}
		\includegraphics[height=5.5 cm]{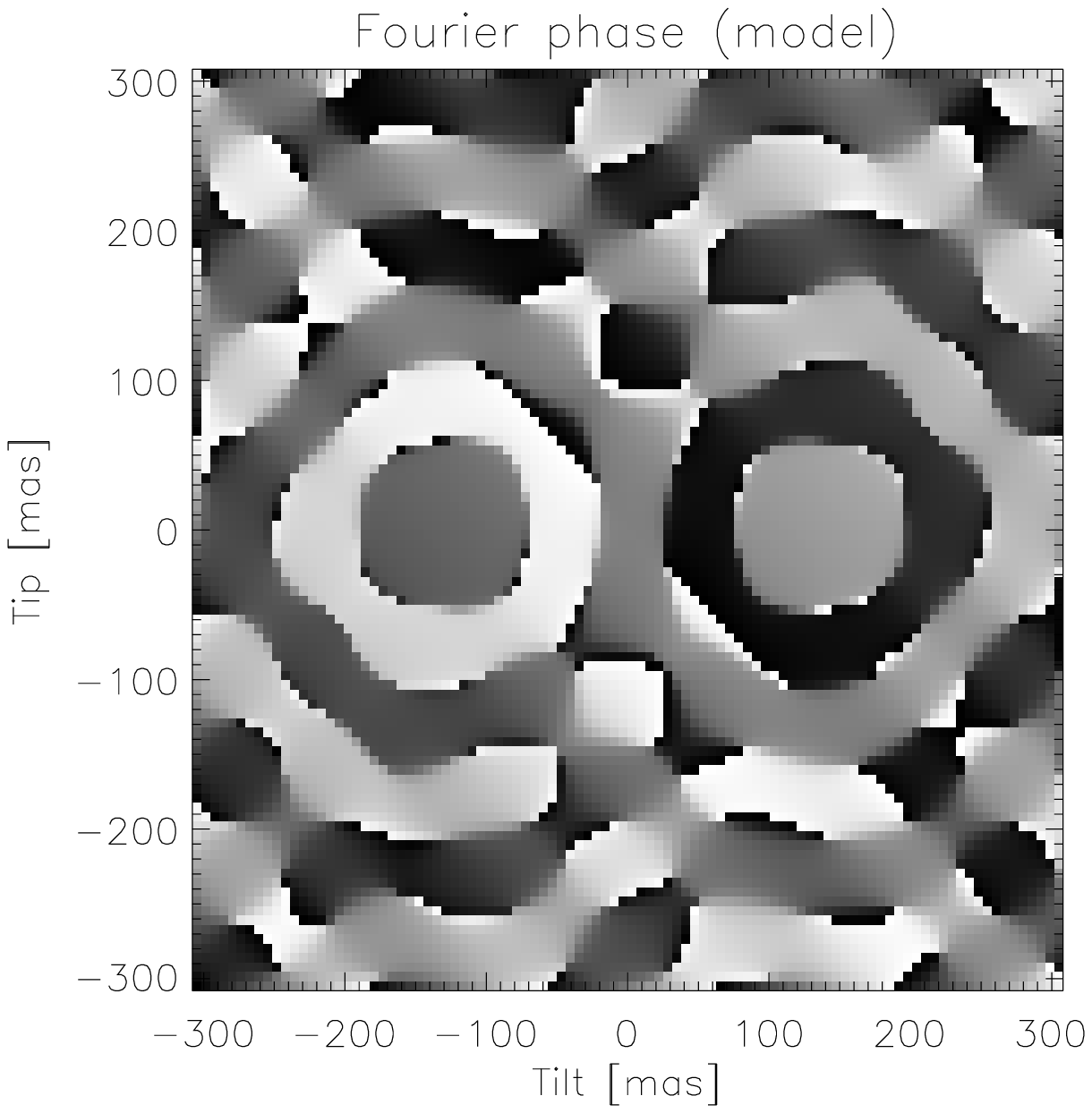}
		\includegraphics[height=5.5 cm]{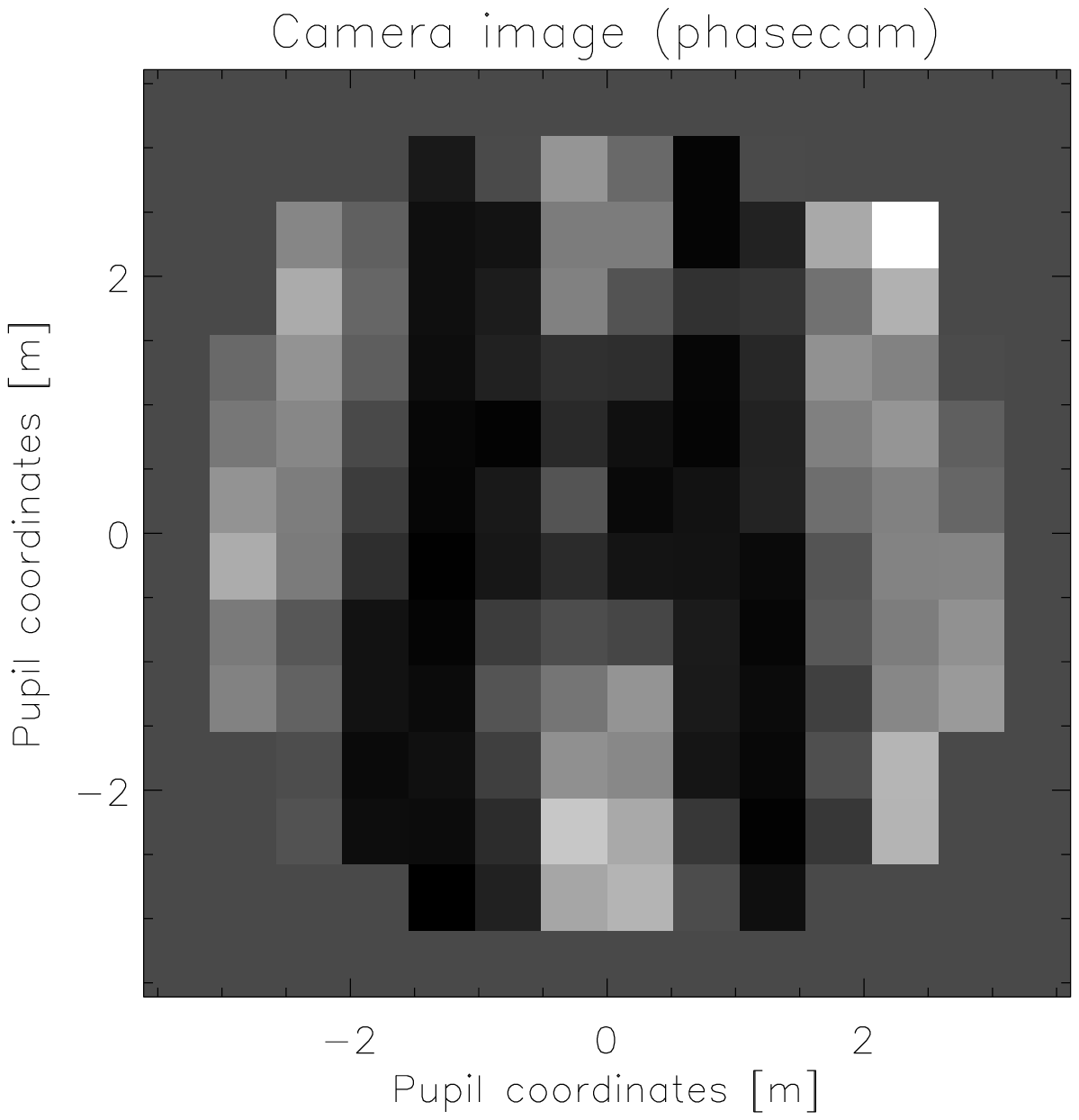}
		\includegraphics[height=5.5 cm]{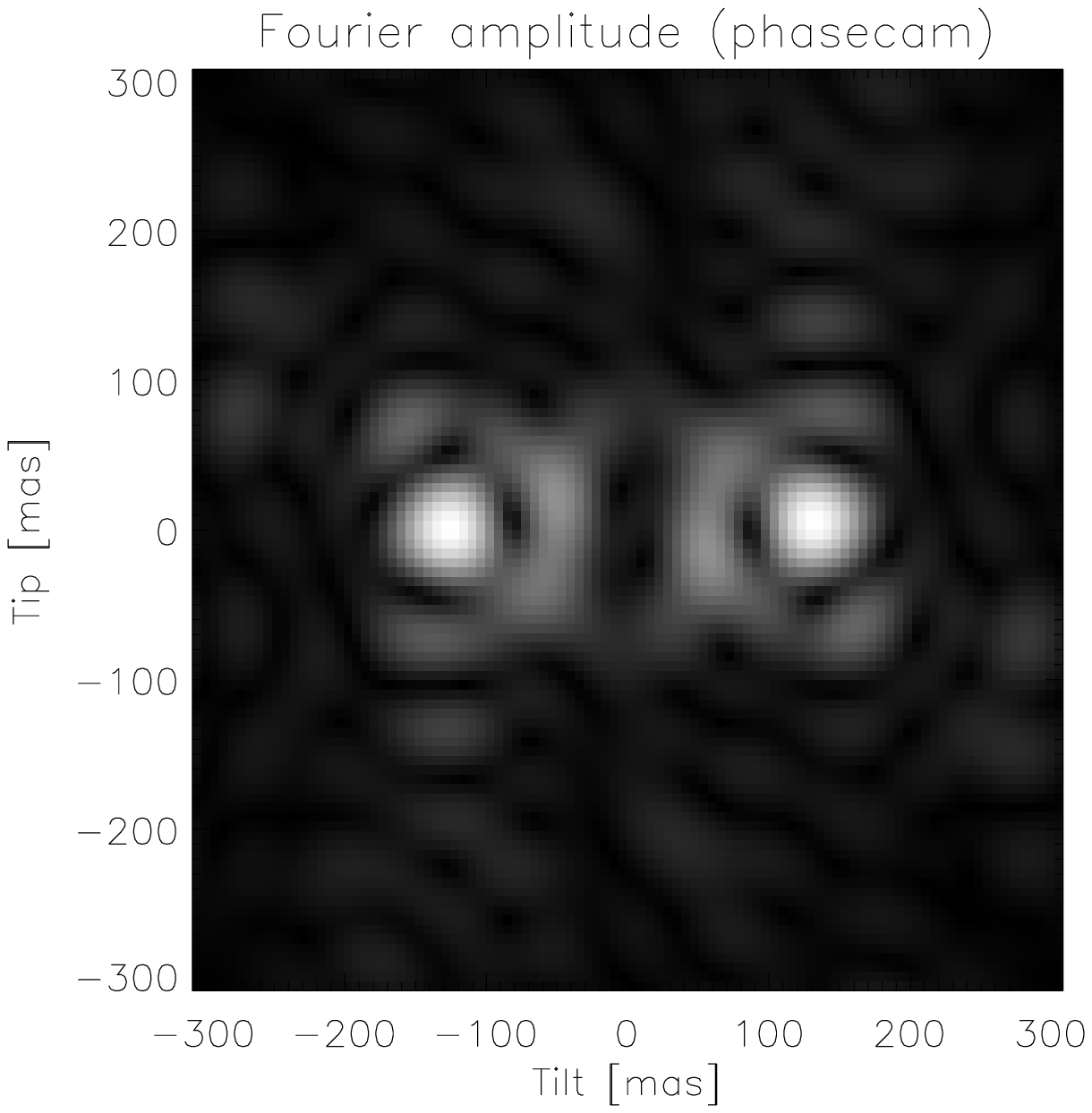}
		\includegraphics[height=5.5 cm]{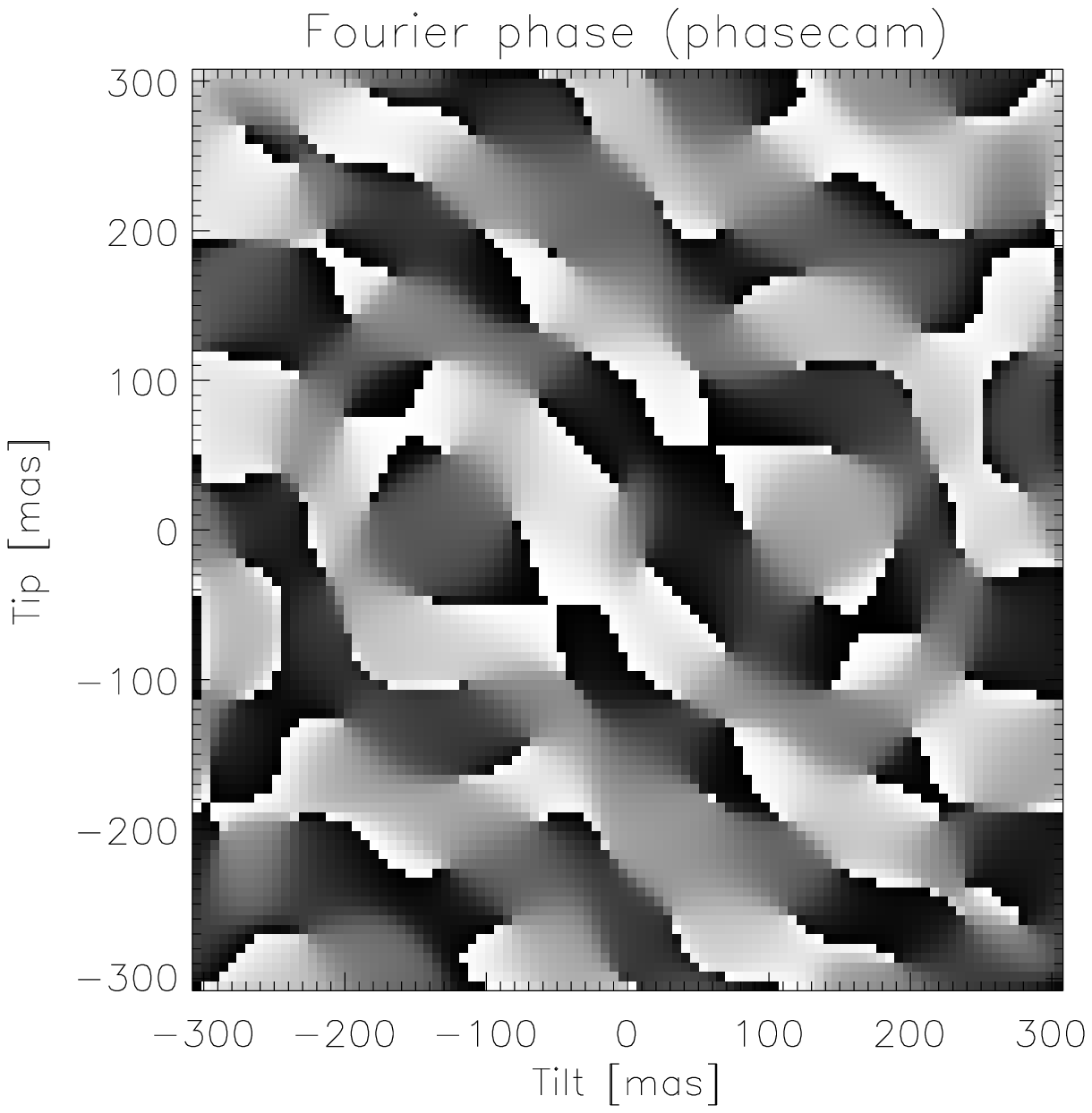}
		\caption{LBTI's phase sensing approach (noise-free model on top and on-sky data from March 17th 2014 on the bottom). Pupil images of two interferometric outputs are formed on PHASECam (one output shown on the left) and the Fourier transform is computed to sense both tip/tilt and phase. The peak position in the amplitude of the Fourier image (middle images) provides the tip/tilt error signal while the argument of the Fourier image (right images) at the peak position provides the phase.} \label{fig:approach}
	\end{center}
\end{figure}

So far, fringe sensing has been mainly performed using an image of pupil fringes (equivalent to wedge fringes). Because of dispersion between 2~$\mu$m and 10~$\mu$m in the beamsplitter, a well overlapped set of images at 10~$\mu$m corresponds to a tilt difference of roughly 3 fringes across the pupil at 2~$\mu$m. This has the nice feature of providing a signal in the Fourier plane well separated from the zero-frequency component and allow us to separate tip-tilt and phase variations via a Fourier transform of the detected light. The magnitude of the Fourier transform gives a measurement of the tip/tilt while the phase of the Fourier transform gives a measurement of the optical path delay. This approach is represented in Figure~\ref{fig:approach} for a noise-free model (top row) and on-sky data from March 17th 2014 (bottom row). Tip/tilt and fringe sensing are currently carried out at 1\,kHz and could go as fast as 2 kHz in the near future. 

\subsection{Group delay}

Because of the nature of the Fourier transform measurement, the phase delay approach described in the previous section is limited to phase values in the  [$-\pi$,$\pi$] range. Any phase error larger than $\pi$
will not be fully detected and, hence, not corrected. On-sky verification showed that such phase jumps occur occasionally even with the phase loop closed at 1\,kHz. In order to get around this issue, the envelope of interference (or the group delay) is tracked simultaneously via the change in contrast of the fringes. A metric called contrast gradient (CG) has been defined as follows:

\begin{equation}
CG = \frac{\sum_i\mid I_i-<I>\mid(x_i-<x>)}{\sum_i I_i};\\
\end{equation}

\noindent where $I_i$ is the intensity for a particular pixel in the pupil and $x_i$ is the coordinate in the horizontal direction of that pixel. The first on-sky stabilized fringes were obtained in December 2013 using this approach and a loop repetition frequency of 1\,kHz. However, despite providing stability, this approach has a limited precision of approximately 1\,$\mu$m and produces a non-Gaussian phase distribution which makes the use of advanced data reduction techniques harder to use. The plan is to rely on the phase delay measurements at full speed and monitor the group delay at a slower speed, typically 20-50\,Hz, to detect and correct any phase jumps. Several merging approaches have been developed recently and will be tested on sky during the next LBTI observing run. 


\section{On-sky performance}\label{sec:perfo}

\begin{figure}[!t]
	\begin{center}
		\includegraphics[height=5.5 cm]{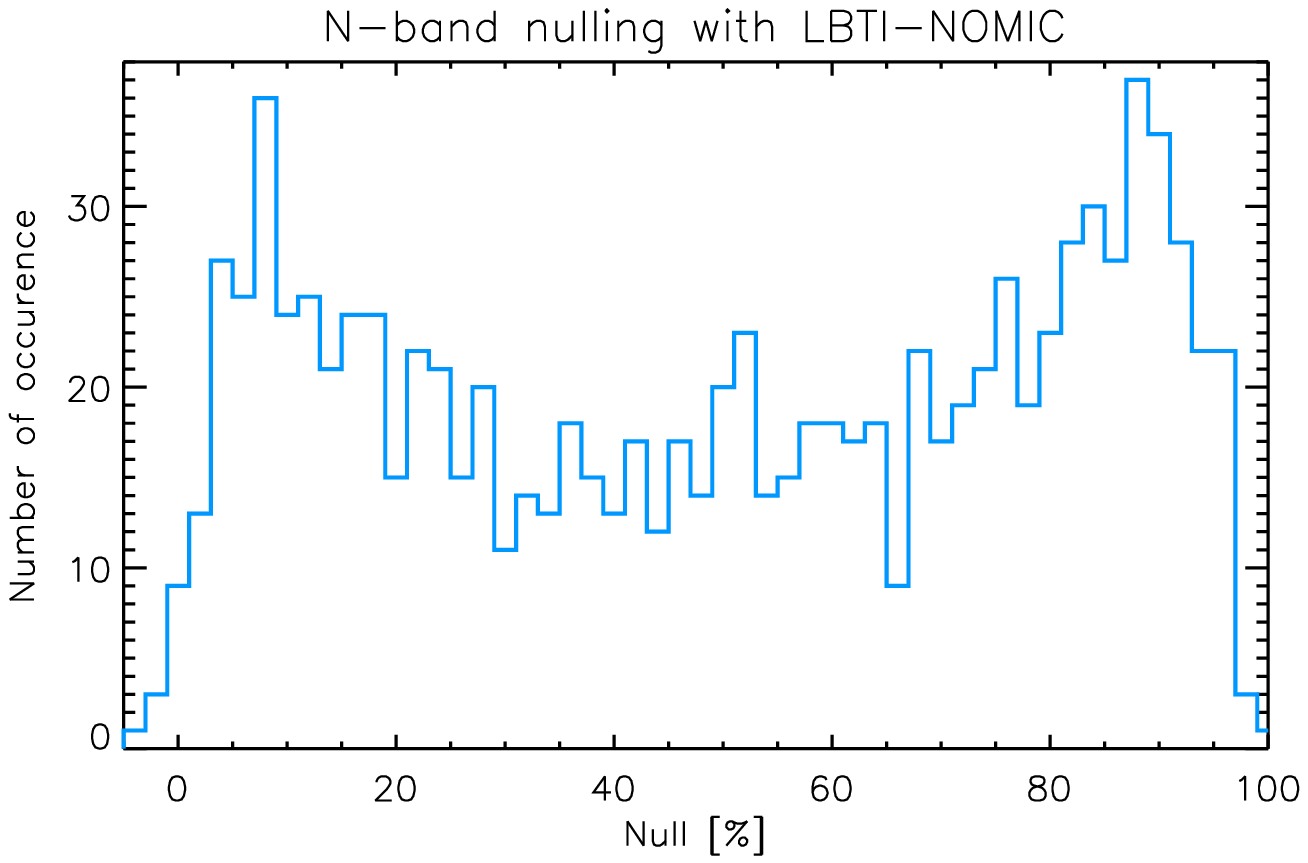}
	         \includegraphics[height=5.5 cm]{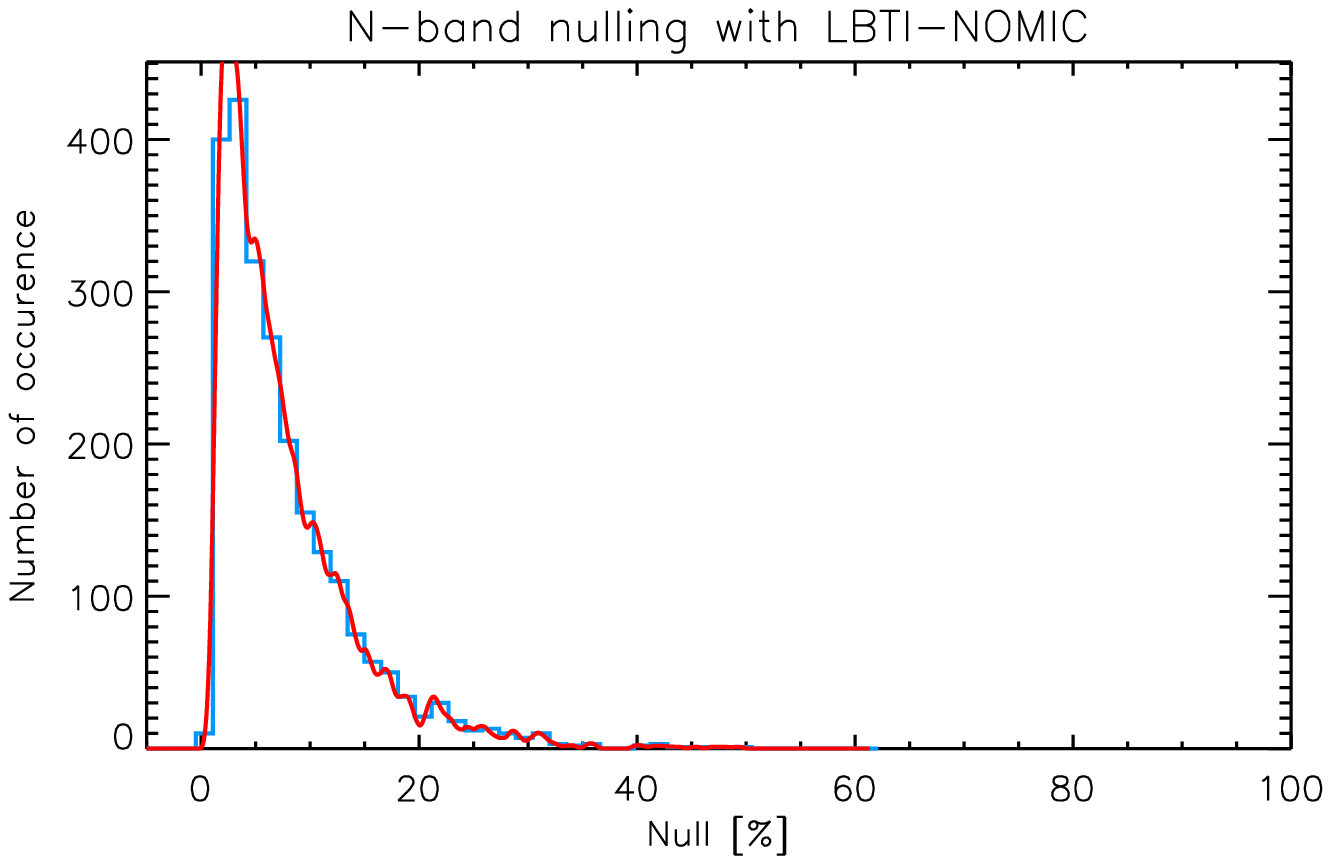}
		\caption{Open-loop null histogram (before December 2013) and closed-loop null histogram (after December 2013) obtained with LBTI/NOMIC. The open-loop null measurements range from near 0 to 100\%, following the random phase variations between the two telescopes, while the closed-loop null measurements are peaking to a null depth of a few percent (as indicated by the red curve representing a Gaussian-kernel fit to the histogram).  The closed-loop null depth is currently limited to a few percent due to an intensity mismatch between the two beams.} \label{fig:null}
	\end{center}
\end{figure}

\begin{figure}[!t]
	\begin{center}
		\includegraphics[height=6.0 cm]{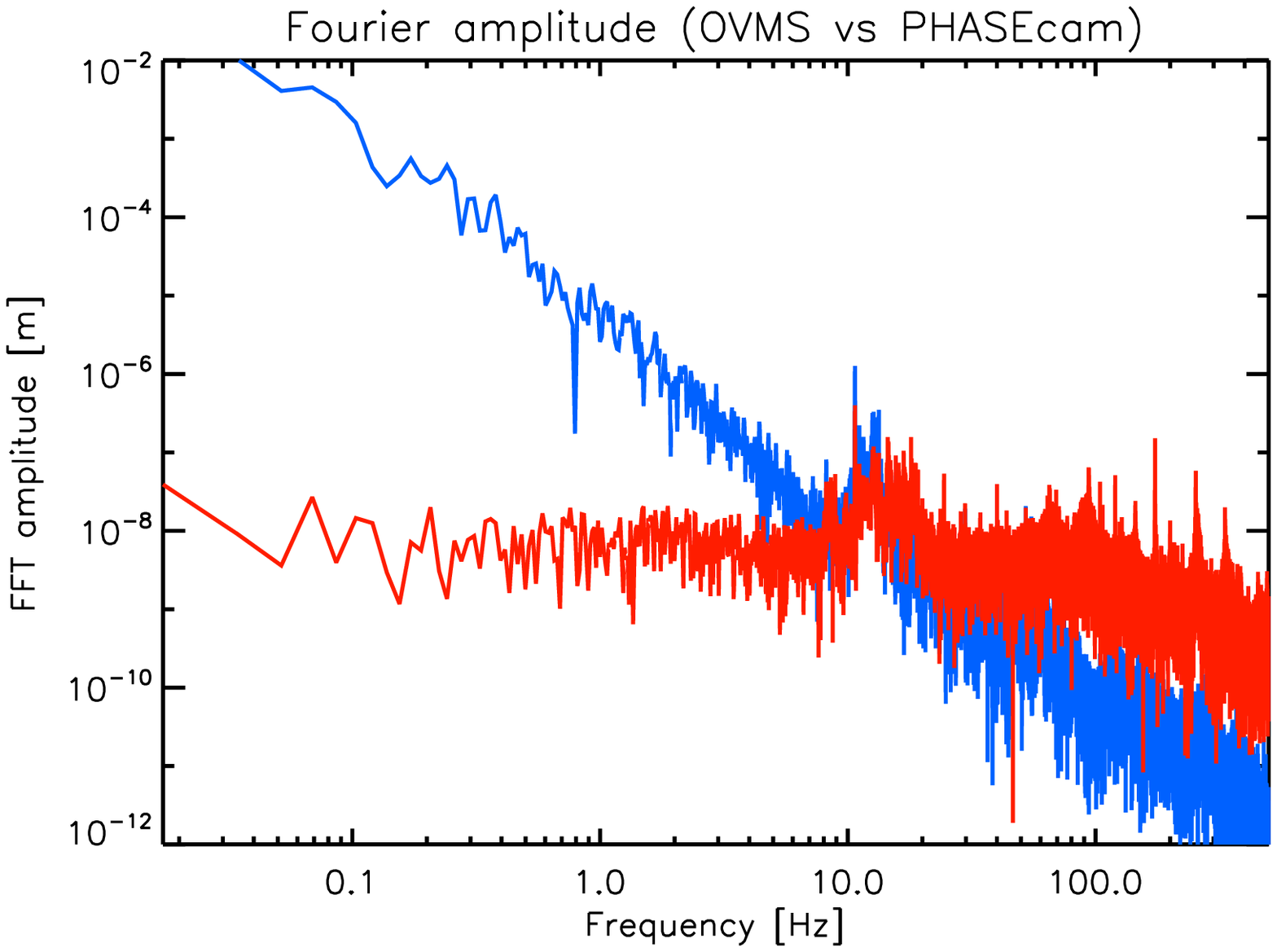}
		\includegraphics[height=6.0 cm]{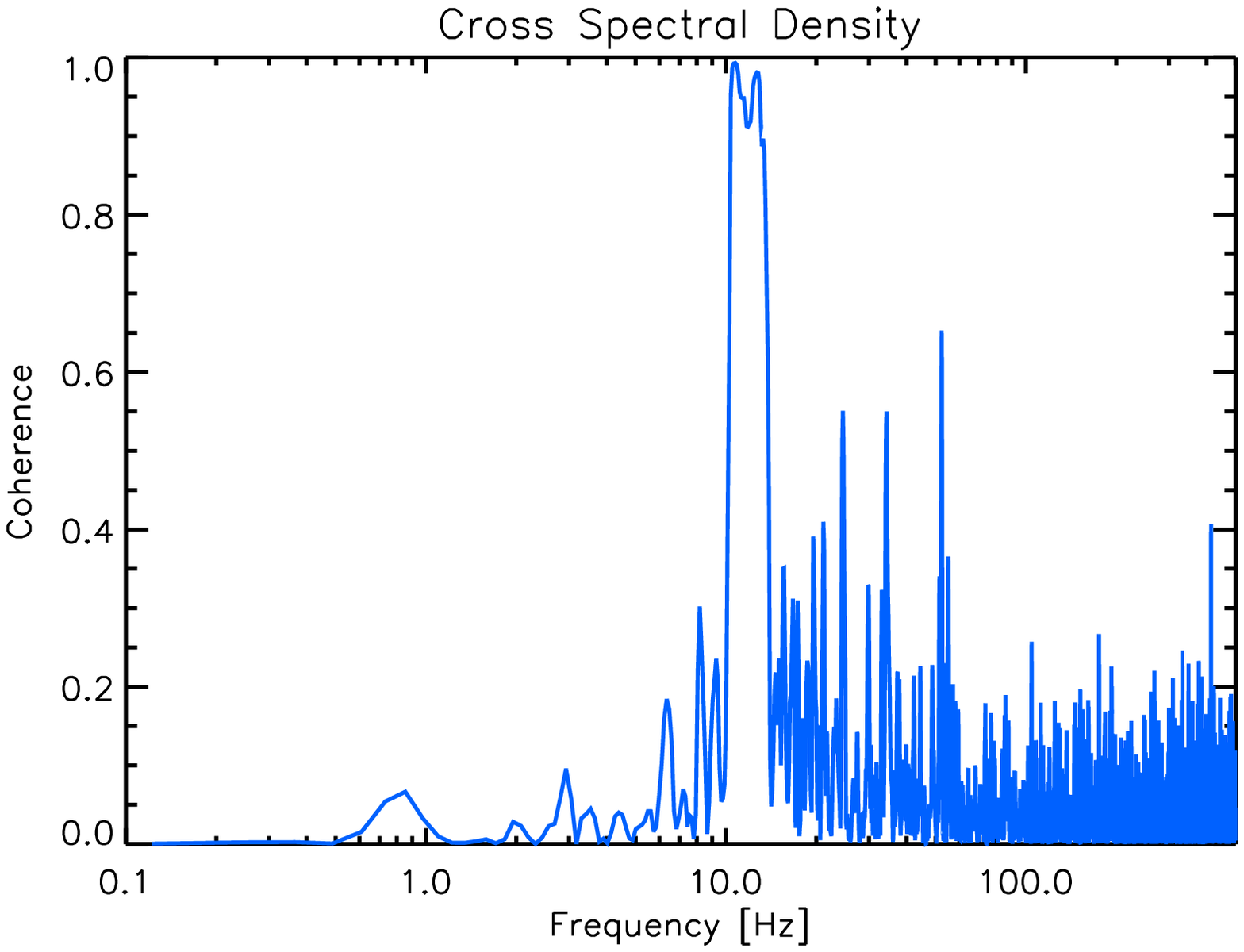}
		\includegraphics[height=6.0 cm]{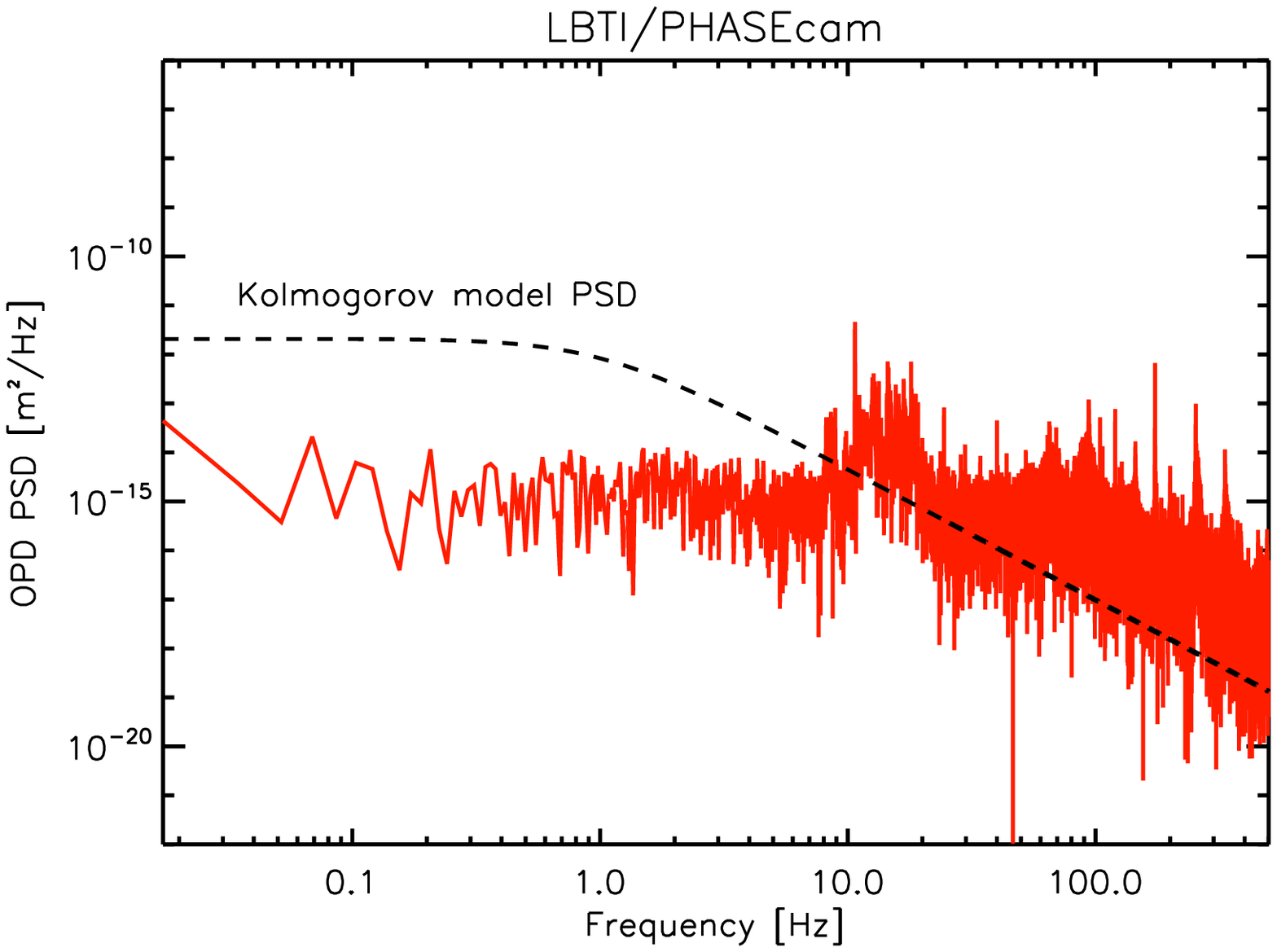}
		\includegraphics[height=6.0 cm]{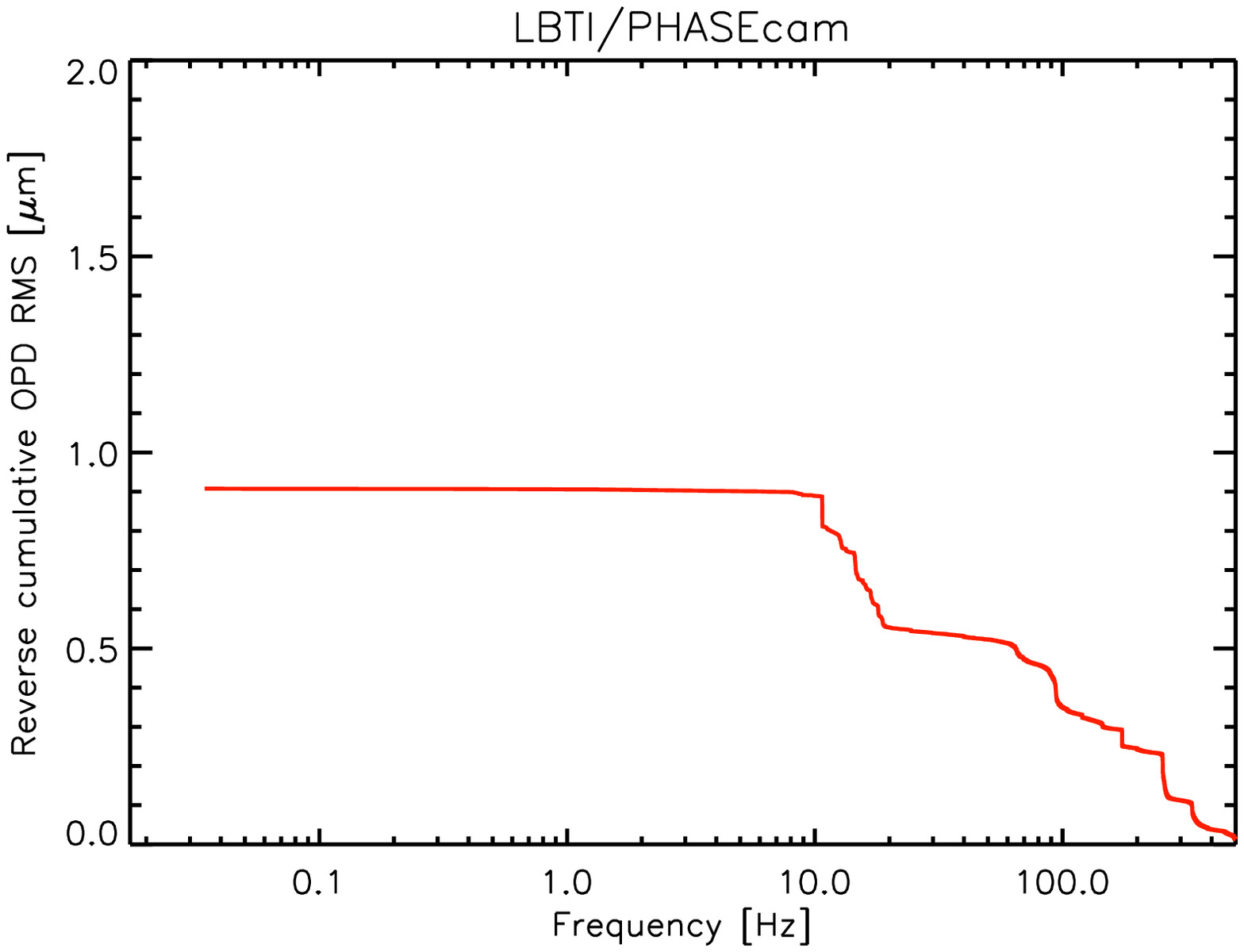}
		\caption{Top left, Fourier spectrum of the differential piston measured by accelerometers located on the two secondary mirrors (blue curve) and corresponding closed-loop Fourier spectrum of the OPD measured by LBTI/PHASECam (red curve). Both data sets have been taken simultaneously on March 11st 2014 at UT5:33. Top right, corresponding cross-spectral density showing good coherence in the 10-13 Hz range. Bottom left, closed-loop power spectral density of the OPD measured by LBTI/PHASECam compared to the power spectral density of a Kolmogorov atmospheric model computed by assuming a Fried parameter of 10\,cm (at 500\,nm), a wind speed of 10\,m/s, and a outer scale of the turbulence of 10\,m (dashed line). Bottom right, corresponding reverse cumulative OPD rms (from 500\,Hz) showing that most of the remaining noise lies in the 10-15\,Hz and 80-500\,Hz frequency ranges.}\label{fig:nic_null}
	\end{center}
\end{figure}

The LBTI has reached several important milestones over the past few years and is on a good track towards routine coherent imaging observations. First fringes were obtained in October 2010, just one month after the installation on the telescope, dual-aperture AO-corrected fringes in April 2012, and first nulling observations in September 2012. The first closed-loop observations were obtained on December 30th 2013 using group delay tracking (see Figure~\ref{fig:null}). The performance of the system was characterized during two following observing runs in February and March 2014 for a range of observing conditions and science object brightnesses. Figure~\ref{fig:nic_null} illustrates the results obtained on March 11, 2014 on the F5V star Procyon (K=-0.65) and a loop repetition frequency of 1\,kHz. The top-left plot shows the Fourier amplitude of the closed-loop OPD measured by PHASECam over a representative 1-min long sequence (red line) and the corresponding differential piston measured simultaneously by accelerometers located on the two secondary mirrors (OVMS \cite{Kurster:2010}, see blue line). The OPD variations are well rejected below a frequency of approximately 80\,Hz with a lot of residual noise in the 10-13 Hz range that is also present on the secondary mirrors.  A formal cross-spectral analysis shows good coherence in this range (see top-right panel) indicating that the residual vibrations detected by PHASECam are well correlated with the differential motion of the secondary mirrors. The closed-loop residual OPD is approximately 900\,nm rms, mainly dominated by noise in the 10-15\,Hz and 80-500\,Hz frequency ranges as shown in the bottom right panel of Figure~\ref{fig:nic_null}. These OPD variations near 13 Hz probably come from excited eigenmodes of the swing arms that support the secondary mirrors. Various mitigation strategies for these vibrations are currently under study (see Section~\ref{sec:prospects}). 

\section{Summary and future work}\label{sec:prospects}

The LBTI demonstrated stabilized path length observations in December 2013 and is currently delivering the first science results using the full resolution of the LBT. So far, path length stabilization was achieved using group delay tracking which is limited to a precision of approximately 1\,$\mu$m rms. Phase delay provides better precision but cannot detect fringe jumps that occur occasionally on sky, even at a loop repetition frequency of 1\,kHz. Several merging approaches have been developed recently and tested with the internal source. They will be tested on sky during the next LBTI observing run. In addition, we have implemented a feed-forward correction algorithm to correct for the vibrations of the secondary mirrors that are responsible for a significant fraction of the residual OPD. Using real-time measurements obtained from accelerometers installed on the secondary mirrors, we plan to inject path length corrections directly into the LBTI via the fast path length corrector. In the longer term, we plan to implement a Linear Quadratic Gaussian control algorithm to correct for variable high-frequency vibrations ($>$100\,Hz). Simulations using real on-sky PHASECam data have shown that this could reduce the closed-loop OPD rms by a factor of 3. Finally, the observatory is also investigating how to directly mitigate the vibrations (e.g., passive or active damping). Commissioning activities for the LBTI will continue in 2014\cite{Hinz:2014} and will be mainly focused on improving the phase stability.


\acknowledgments     
The authors are grateful to M.~Colavita for his advice. LBTI is funded by a NASA grant in support of the Exoplanet Exploration Program (NSF 0705296). The LBT is an international collaboration among institutions in the United States, Italy and Germany. LBT Corporation partners are: The University of Arizona on behalf of the Arizona university system; Istituto Nazionale di Astrofisica, Italy; LBT Beteiligungsgesellschaft, Germany, representing the Max-Planck Society, the Astrophysical Institute Potsdam, and Heidelberg University; The Ohio State University, and The Research Corporation, on behalf of The University of Notre Dame, University of Minnesota and University of Virginia. 

\bibliographystyle{spiebib}   




\end{document}